\documentclass[aps,pra,twocolumn,superscriptaddress]{revtex4-1}

\usepackage{color}
\usepackage{graphicx}% Include figure files
\usepackage{dcolumn}% Align table columns on decimal point
\usepackage{bm}% bold math
\usepackage{amsmath}
\usepackage{amssymb}
\usepackage{latexsym}
\usepackage{epsfig}
\usepackage{amsbsy}
\usepackage{array}
\usepackage{amssymb}
\usepackage{setspace}
\usepackage{bm}
\usepackage{indentfirst}
\usepackage{subfigure}
\usepackage{float}
\usepackage{booktabs}
\usepackage{threeparttable}
\usepackage{tablefootnote}

\def\sint{\ifmmode{- \!\!\!\!\!\! \int}
    \else{\hbox{$- \!\!\!\! \int \ $}}\fi}

\newcommand{\Rmnum}[1]{\uppercase\expandafter{\romannumberal #1}}

\begin{document}

\preprint{pra}

\title{On the theory of solid-state harmonic generation governed by crystal symmetry}

\author{Chen Qian} \affiliation{Institute of Ultrafast Optical Physics, MIIT Key Laboratory of Semiconductor Microstructure and Quantum Sensing, Department of Applied Physics, Nanjing University of Science and Technology, Nanjing 210094, P R China} \affiliation{Institute of Physics, Chinese Academy of Sciences, Beijing National Laboratory for Condensed Matter Physics, Beijing 100190, P R China}
\author{Shicheng Jiang}\affiliation{State Key Laboratory of Precision Spectroscopy, East China Normal University, Shanghai 200062, China}
\author{Tong Wu} \affiliation{Institute of Ultrafast Optical Physics, MIIT Key Laboratory of Semiconductor Microstructure and Quantum Sensing, Department of Applied Physics, Nanjing University of Science and Technology, Nanjing 210094, P R China}
\author{Hongming Weng}\email[ ]{hmweng@iphy.ac.cn}\affiliation{Institute of Physics, Chinese Academy of Sciences, Beijing National Laboratory for Condensed Matter Physics, Beijing 100190, P R China} \affiliation{School of Physics, University of Chinese Academy of Sciences, Beijing 100049, China} \affiliation{Songshan Lake Materials Laboratory, Dongguan, Guangdong 523808, China}
\author{Chao Yu}\email[ ]{chaoyu@njust.edu.cn}\affiliation{Institute of Ultrafast Optical Physics, MIIT Key Laboratory of Semiconductor Microstructure and Quantum Sensing, Department of Applied Physics, Nanjing University of Science and Technology, Nanjing 210094, P R China}
\author{Ruifeng Lu}\email[ ]{rflu@njust.edu.cn}\affiliation{Institute of Ultrafast Optical Physics, MIIT Key Laboratory of Semiconductor Microstructure and Quantum Sensing, Department of Applied Physics, Nanjing University of Science and Technology, Nanjing 210094, P R China}

\date{\today}% It is always \today, today,

\begin{abstract}
The solid-state harmonic generation (SSHG) derives from photocurrent coherence. The crystal symmetry, including point-group symmetry and time-reversal symmetry, constrains the amplitude and phase of the photocurrent, thus manipulates the coherent processes in SSHG. We revisit the expression of photocurrent under the electric dipole approximation and give an unambiguous picture of non-equilibrium dynamics of photocarriers on laser-dressed effective bands. In addition to the dynamical phase, we reveal the indispensable roles of the phase difference of transition dipole moments and the phase induced by shift vector in the photocurrent coherence. Microscopic mechanism of the selection rule, orientation dependence, polarization characteristics, time-frequency analysis and ellipticity dependence of harmonics governed by symmetries is uniformly clarified in our theoretical framework. This work integrates non-equilibrium electronic dynamics of condensed matter in strong laser fields, and paves a way to explore more nonlinear optical phenomena governed by crystal symmetry.
\\ \par\

\end{abstract}

\maketitle
%\tableofcontents
\section{Introduction }
A large number of statistically significant photons exhibit strong wave property, and many phenomena in strong-field physics can be attributed to the interference of light waves. The nonlinear photocurrent in a material driven by strong laser fields can coherently emit discrete solid-state harmonic generation (SSHG) \cite{1,2}. Since the dynamical process of accelerated carrier is very sensitive to intrinsic properties of materials, SSHG has the capability of detecting the band structure \cite{3,4,5}, topological geometries \cite{6,7,8,9,10,11,12,13,14} and strongly correlated interaction \cite{15,16,17,18}. Each of these relates to the knowledge of crystal symmetry. The space-time symmetry of the applied field combined with the crystal symmetry provides strict coherence conditions for photocurrent, which can be recorded by harmonic signal that conforms to the selection rules \cite{19,20,21,22}. In the last decade, the correspondence between crystal symmetry and SSHG has been confirmed adequately in literatures. However, a complete microscopic framework for illuminating the photocurrent coherence in solids has not yet been established.

 In the absence of external fields, the symmetry of band structures and wave functions is prescribed by the crystal symmetry. With the addition of ultrafast oscillating laser fields, a handful of electrons are excited into the conduction band and form paired electric dipoles with holes in the valence band. The dipoles are forced to oscillate in the electric potential formed by the Coulomb and laser fields, coherently producing harmonic radiation. Much intrinsic information about crystal bands can be traced by these moving dipoles. 

The roles of the band structure \cite{23}, Berry curvature \cite{7}, transition dipole moment \cite{24}, and shift vector \cite{14} have been successively reported to reveal the dynamical process of photoelectrons. These theoretical explorations are consistent with the experimental results but have not yet formed a systematic thinking. For an example, even-order harmonics in the vertical polarization can be induced by the Berry curvature, group velocity, or interband transition as previously reported \cite{7,23,24}. However, the Berry curvature comes from first-order correction in the electron transition approximated by the perturbation theory, and its contribution can be attributed to the interband transition process when non-perturbative transition dominates \cite{6,25}. Therefore, the origin of harmonics should be reviewed based on the behavior of electron transitions. In addition, the theoretical analysis proposed by Vampa \emph{et al}. in 2014 shows that the phase of photocurrent only contains the dynamical phase \cite{26}. Moreover, Jiang \emph{et al}. emphasize the indispensability of the transition dipole phase (TDP) that it is unreasonable to artificially remove it during numerical calculations \cite{24,27}. Later, Li \emph{et al}. and Yue \emph{et al}. confirm that the Berry connection should also be considered to ensure the gauge invariance of the photocurrent \cite{28,29}. In recent years, the numerical method for calculating time-dependent photocurrent using density matrix equations has been improved, which is qualitatively consistent with experiments \cite{30,31,32,33,34}. Nevertheless, the microscopic interference process of photocurrent involved has not been clarified. In particular, the effects of shift vector and TDP are poorly understood in strong-field physics. 

In this paper, we derive the selection rule of SSHG via analytic expressions of photocurrent and reveal the role of TDP difference in the photocurrent coherence. We are committed to clarifying the symmetry dependence of SSHG with a fundamental picture involving the TDP difference and shift vector. In particular, we will use our theoretical framework to discuss the orientation dependence, polarization property, time-frequency analysis, and ellipticity dependence of harmonics determined by crystal symmetry.

\section{Theory}

Considering the rationality of single-electron and dipole approximations in appropriate strong-field environment, we derive expressions of interband and intraband currents through two-band semiconductor Bloch equations as follows (see Appendix A for details, atomic units are used throughout unless otherwise stated): 
\begin{widetext}
\begin{equation}
\begin{aligned}
\mathbf{J}_{n m}^a(t)=&-\frac{1}{N_{\mathrm{c}}} \sum_{\mathbf{K} \in \mathrm{BZ}} \int_{-\infty}^t d t^{\prime} \varepsilon_{n m}(\mathbf{k}(t))\left|\mathbf{d}_{n m}^a(\mathbf{k}(t))\right|\left[\mathbf{E}^b\left(t^{\prime}\right)  \cdot\left|\mathbf{d}_{m n}^b\left(\mathbf{k}\left(t^{\prime}\right)\right)\right|\right] f_{n m}(\mathbf{K}, t) \\
&\times e^{-i\left[S_{\mathrm{dyn}}\left(\mathbf{K}, t, t^{\prime}\right)+S_{\mathrm{shift}}\left(\mathbf{K}, t, t^{\prime}\right)+S_{\Delta \mathrm{TDP}}(\mathbf{k}(t))\right]},
\end{aligned}
\end{equation}
\begin{equation}
\begin{aligned}
\mathbf{J}_{n n}^a(t)=&\frac{1}{N_{\mathrm{c}}} \sum_{\mathbf{K} \in \mathrm{BZ}} \int_{-\infty}^t d t^{\prime} \int_{-\infty}^{t^{\prime}} d t^{\prime \prime} \partial_{\mathbf{k}_a} \varepsilon_n(\mathbf{k}(t))\left[\mathbf{E}^b\left(t^{\prime}\right) \cdot\left|\mathbf{d}_{n m}^b\left(\mathbf{k}\left(t^{\prime}\right)\right)\right|\right] 
{\left[\mathbf{E}^b\left(t^{\prime \prime}\right) \cdot\left|\mathbf{d}_{m n}^b\left(\mathbf{k}\left(t^{\prime \prime}\right)\right)\right|\right] }f_{n m}\left(\mathbf{K}, t^{\prime}\right)\\
&\times e^{-i\left[S_{\mathrm{dyn}}\left(\mathbf{K}, t^{\prime}, t^{\prime \prime}\right)+S_{\mathrm{shift}}\left(\mathbf{K}, t^{\prime}, t^{\prime \prime}\right)+S_{\Delta \mathrm{TDP}}\left(\mathbf{k}\left(t^{\prime}\right)\right)\right]}+\mathrm{c.c.},
\end{aligned}
\end{equation}
\end{widetext}
where $a$ and $b$ are Cartesian indices, labeling the directions of the currents $\mathbf{J}(t)$ and the electric field $\mathbf{E}(t)$, respectively. ${N_{\mathrm{c}}}$ is the total number of unit cells. Houston basis is used here. The quasi momentum $\mathbf{k}(t)$ of electrons changes adiabatically with the laser field, and the evolution relationship is $\mathbf{k}(t)=\mathbf{K}+\mathbf{A}(t)$. $\mathbf{K}$ is the canonical momentum of the crystal in the absence of field. Here, the transition dipole matrix element 
$\mathbf{d}_{n m}(\mathbf{k})=i\left\langle u_{n, \mathbf{k}}\left|\partial_{\mathbf{k}}\right| u_{m, \mathbf{k}}\right\rangle$ describes the polarization of electron-hole pairs. $f_{n m}(\mathbf{K}, t)=\rho_{n n}(\mathbf{K}, t)-\rho_{m m}(\mathbf{K}, t)$ is the difference of Fermi-Dirac distribution, and the band index $n\neq m$.

There are three phase factors that determine the photocurrent coherence. Firstly, the dynamical phase
\begin{equation}
\begin{aligned}
S_{\mathrm{dyn}}\left(\mathbf{K}, t, t^{\prime}\right)=\int_{t^{\prime}}^t \varepsilon_{m n}(\mathbf{k}(\tau)) d \tau,
\end{aligned}
\end{equation}
where $\varepsilon_{m n}(\mathbf{k})=\varepsilon_m(\mathbf{k})-\varepsilon_n(\mathbf{k})$ is the energy difference between bands $n$ and $m$.

Secondly, a shift phase is introduced as
\begin{equation}
\begin{aligned}
S_{\mathrm{shift}}\left(\mathbf{K}, t, t^{\prime}\right)=\int_{t^{\prime}}^t \mathbf{E}^b(\tau) \cdot \mathbf{R}_{m n}^{b, b}(\mathbf{k}(\tau)) d \tau,
\end{aligned}
\end{equation}
where the shift vector $\mathbf{R}_{m n}^{b, b}(\mathbf{k})=\mathbf{d}_{m m}^b(\mathbf{k})-\mathbf{d}_{n n}^b(\mathbf{k})-\partial_{\mathbf{k}_b} \phi_{m n}^b(\mathbf{k})$, formed by the Berry connections and TDP, represents the offset of charge centers of different bands \cite{14,35}. $\phi_{m n}^a(\mathbf{k})$ is the TDP as $\mathbf{d}_{m n}^a(\mathbf{k})=\left|\mathbf{d}_{m n}^a(\mathbf{k})\right| e^{i \phi_{m n}^a(\mathbf{k})}$. $\left|\mathbf{d}_{m n}^a(\mathbf{k})\right|$ is the transition dipole amplitude and denotes the polarization intensity of electron-hole pair.

The third phase factor
\begin{equation}
\begin{aligned}
S_{\Delta \mathrm{TDP}}(\mathbf{k})=\phi_{m n}^a(\mathbf{k})-\phi_{m n}^b(\mathbf{k})
\end{aligned}
\end{equation}
denotes the difference of transition dipole phases ($\Delta \mathrm{TDP}$) along $a$ and $b$ directions, which comes from the deflection of non-collinear currents relative to the driving field.

Each of the three phase factors is gauge independent. The total current is $\mathbf{J}^a=\sum_{n, m}\left(\mathbf{J}_{n m}^a+\mathbf{J}_{n n}^a\right)$. From Eqs. (1) and (2), we notice that the phases of interband and intraband currents have the same form; thus, the symmetry dependence of their coherence in laser-crystal systems is always consistent.

\subsection{$\Delta \mathbf{TDP}$-Determined Selection Rules of SSHG }

One of the most widely studied and robust laws in SSHG is its selection rule, which mainly depends on the crystal symmetry and can be divided into two categories. The first type appears in the typical Floquet systems and results from periodic oscillations of laser fields. Photoexcited carriers can display dynamical symmetry and coherently generate harmonic radiation \cite{22}. The other is induced directly by the crystal symmetry that is not broken by applied fields. The photocurrents cancel each other out, leading to no harmonic in some particular directions. 

In the strong-laser region, external electric field can be compared with the coulomb field in crystals, it cannot be regarded as a perturbation. In this case, we need to consider the influence of the time-dependent population of charge density on the nonlinear process. Therefore, compared with the perturbation approximation, a broader theory is expected to treat systems under strong laser fields. 

The crystal symmetry we considered here includes the point-group symmetry and time-reversal symmetry. Due to the periodic translational symmetry, crystals are limited to 32 point groups and 122 magnetic point groups. In addition, the laser fields could also contain abundant time-space symmetry. By combining the symmetries of crystals and light fields, we can obtain a wide variety of selection rules for SSHG. In this paper, we further explore more fundamental microscopic dynamics underlying these rules.

Consider a point-group symmetry operation $\hat{G}$ on the transition dipole matrix element (see Appendix B for derivations), we have
\begin{equation}
\begin{aligned}
\hat{G} \boldsymbol{d}_{n m}^a(\boldsymbol{k}) \hat{G}^{\dagger}=\boldsymbol{d}_{n m}^{a^{\prime}}\left(G^{-1} \boldsymbol{k}\right)=G_{a^{\prime} a} \boldsymbol{d}_{n m}^a(\boldsymbol{k}),
\end{aligned}
\end{equation}
in which $a, a^{\prime}=G^{-1} a$ denote the directions of the transition dipole moments, $G_{a^{\prime} a}=e^{i\left[\phi_{n m}^{a^{\prime}}(\mathbf{k}(t))-\phi_{n m}^a(\mathbf{k}(t))\right]}$ is just the $\Delta \mathrm{TDP}$ between $a$ and $a^{\prime}$ direction under $\hat{G}$. Moreover, the scalar quantities such as the dynamical phase and shift phase are invariant under $\hat{G}$.

With the addition of the light field, let's combine $\hat{G}$ and order-$N$ time translation operator ($N \in \mathbb{N}$, $\mathbb{N}$ denotes set of natural numbers):
\begin{equation}
\begin{aligned}
\hat{X}=\hat{G} \cdot \hat{\tau}_N
\end{aligned}
\end{equation}
with $\hat{\tau}_N t \equiv t \pm \frac{T_0}{N}, T_0=\frac{2 \pi}{\omega_0}$ is the period of the laser field of frequency $\omega_0$. We apply the dynamical symmetry operation to the photocurrent,
\begin{equation}
\begin{aligned}
\hat{X} \mathbf{J}^a(\mathbf{K}, t) \hat{X}^{\dagger}=\mathbf{J}^{G^{-1} a}\left(G^{-1} \mathbf{K}, \hat{\tau}_N t\right).
\end{aligned}
\end{equation}
Every wave vector $\mathbf{K}$ in the lattice is a candidate for harmonic peaks unless symmetry forbids it. For the time-dependent quasi momentum $\mathbf{k}(t) \equiv \mathbf{K}+A(t)$,
\begin{equation}
\begin{aligned}
\hat{X} \mathbf{k}(t)=G^{-1} \mathbf{K}+\mathbf{A}\left(\hat{\tau}_N t\right).
\end{aligned}
\end{equation}
The action of the laser field may disrupt initial symmetries of crystals, but new dynamical symmetry can be induced. Based on above transformation rules, we can derive that the interband current with dynamical symmetry can transform as
\begin{equation}
\begin{aligned}
\begin{aligned}
\hat{X} \mathbf{J}_{n m}^a(\mathbf{K}, t) \hat{X}^{\dagger} & =\mathbf{J}_{n m}^{G^{-1} a}\left(G^{-1} \mathbf{K}, \hat{\tau}_N t\right) \\
& =-i \varepsilon_{n m}(\mathbf{k}(t)) G_{a^{\prime} a} \mathbf{d}_{n m}^a(\mathbf{k}(t)) \rho_{n m}(\mathbf{K}, t) \\
& =\mathbf{J}_{n m}^a(\mathbf{K}, t) e^{i S_{\Delta \mathrm{TDP}}(\mathbf{k}(t))}.
\end{aligned}
\end{aligned}
\end{equation}
The electron density $\rho_{n m}$ is approximatively invarible under the dynamical symmetry operation. Same transformation rule is followed for the intraband counterpart. 

\begin{figure}
    \includegraphics[width=3.4in,angle=0]{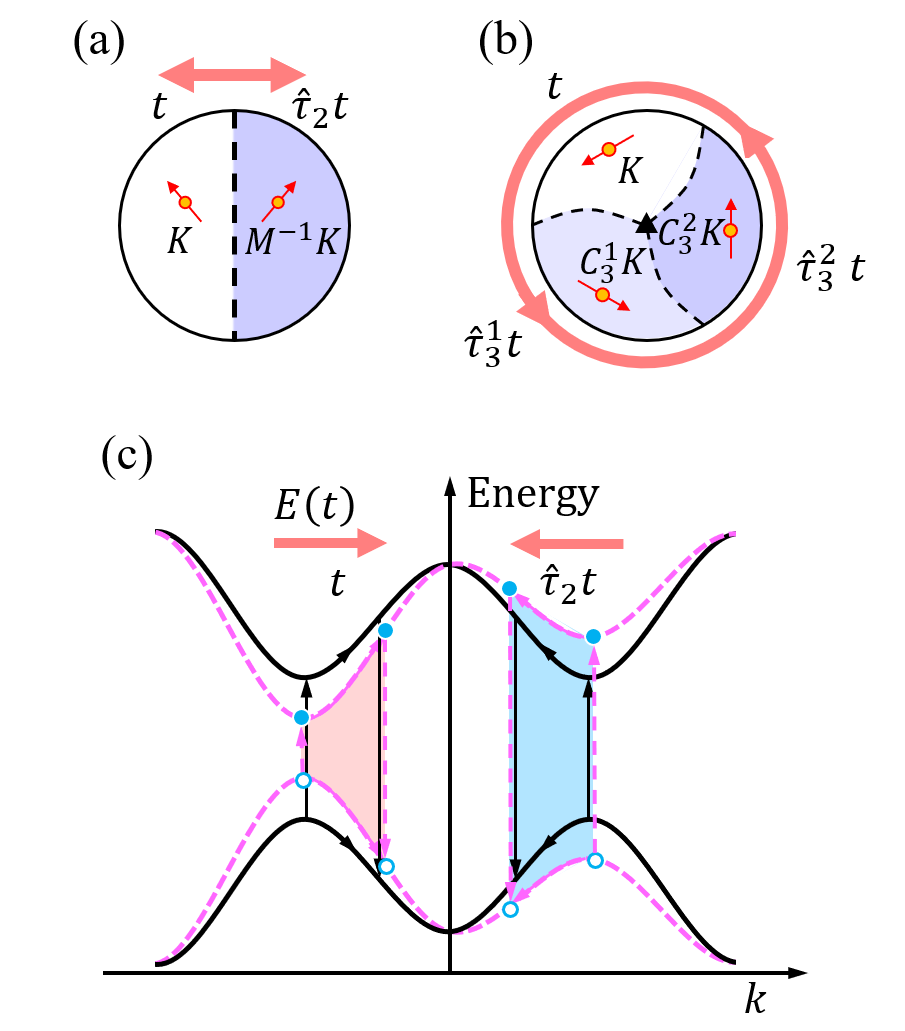}
    \caption{Diagrams of dipole interference induced by dynamical symmetries. (a) mirror symmetry combined with order-2 temporal symmetry of light. (b) 3-fold rotational symmetry combined with order-3 temporal symmetry of light. The red arrows represent the polarization of laser field. The dipoles excited in different subcycles are indicated by dotted arrows, which are linked by corresponding symmetry operations, respectively. (c) Interference pattern of two channels in reciprocal space. The solid black lines are symmetric bands and channels protected by the inversion symmetry. The magenta dashed line is the effective bands deformed by the laser field, its asymmetry comes from the inversion symmetry breaking of crystals. The arrows between the bands indicate electronic excitation and electron-hole recombination, while the arrows along the band dispersions show acceleration of electron-hole pair. Different color areas surrounded by the arrows highlight the asymmetry of the channels.}\label{Fig_1}
\end{figure}

We find that all the point-group dynamical symmetry operations only induce a change of photocurrent phase, which is the $\Delta \mathrm{TDP}$. That is, the transformation occurs only at the argument or the phase of the transition dipole, but the band dispersion, transition dipole amplitude, dynamical phase and shift phase are invariant. The coherence of harmonics stem from the interference between transition dipoles with different arguments associated by the dynamic symmetry (Fig. 1(a) and 1(b) show schematics of coherent dipoles under the mirror symmetry and rotational symmetry, respectively). More detailed derivation for typical cases of basic symmetry operations can be found in Appendix C. In Table I and Table II, $\Delta \mathrm{TDP}$-determined selection rules of SSHG are shown for the mirror symmetry and rotational symmetry.

\begin{table}[t!]
\centering
\caption{Selection rule of SSHG by mirror symmetry}
\resizebox{8.7cm}{7.7mm}{
\begin{tabular}{ccc}
\hline
\hline
 & $a \parallel c$ & $a \perp c$ \\
\hline

$b \parallel c$ & $\Delta \mathrm{TDP}=\pi, \omega=(2 l+1) \omega_0$ & $\Delta \mathrm{TDP}=0, \omega=2 l \omega_0$ \\
$b \perp c$ & no harmonics & integer order harmonics \\
\bottomrule
\hline
\hline
\end{tabular}}\\

\footnotesize{$a, b$ represent directions of photocurrent, laser field, respectively, and $c$ is the normal direction of crystal mirror plane. $l \in \mathbb{N}$}

\end{table}

\begin{table}[t!]

\centering
\caption{Selection rule of SSHG by rotational symmetry}

\resizebox{8.7cm}{7.7mm}{

\begin{tabular}{ccc}
\hline
\hline
 & $a \perp c$ & $a \parallel c$ \\
\hline

$b \perp c$ & $\Delta \mathrm{TDP}= \pm \frac{2 \pi}{N}, \omega=(N l \pm 1) \omega_0$ & $\Delta \mathrm{TDP}=0, \omega=N l \omega_0$ \\
$b \parallel c$ & no harmonics & integer order harmonics \\
\bottomrule
\hline
\hline
\end{tabular}}\\
\footnotesize{$c$ is the direction of crystal rotational axis, the other parameters are the same as Table I.}
\end{table}

\subsection{Role of Shift Phase in Three-Step Model of SSHG}

In previous works of strong-field physics, similar to the expression of gas-state harmonics under strong-field approximation, only the dynamical phase induced by coulomb barriers has been considered. However, the role of shift phase induced by applied electric barriers in SSHG has not been clarified so far. To demonstrate indispensable role of shift phase in the process of photocurrent coherence, we compare order-2 dynamical coherence processes formed by spatial-inversion symmetry with time-reversal symmetry. 

The second harmonic generation is one of the most common methods to determine the inversion symmetry of crystals. We know that when we apply a monochromatic light to a centrosymmetric crystal, even-order harmonics can be canceled by destructive interference, while the odd-order harmonics show constructive interference. 

Let $\hat{P}=\hat{I} \cdot \hat{\tau}_2, \hat{I}$ is the spatial-inversion operation. Using Eq. (10), the interband current under $\hat{P}$ transforms as
\begin{equation}
\begin{aligned}
\hat{P} \mathbf{J}_{n m}^a(\mathbf{K}, t) \hat{P}^{\dagger}=\mathbf{J}_{n m}^{I^{-1} a}\left(-\mathbf{K}, \hat{\tau}_2 t\right)=-\mathbf{J}_{n m}^a(\mathbf{K}, t) .
\end{aligned}
\end{equation}
Integrating over the entire $\mathrm{BZ}$, we then have $\mathbf{J}_{n m}^{I^{-1} a}\left(\hat{\tau}_2 t\right)=-\mathbf{J}_{n m}^a(t)$. The reversal of the current comes from the inversed dipole moment. After applying Fourier transform, we know that the radiated photon frequency is $\omega=(2 l+1) \omega_0, l \in \mathbb{N}$. The same conclusion can be reached by analyzing the intraband current.

Similarly, we define an operator $\hat{U}$ that performs time-reversal operation $\hat{T}$ on the crystals and order-2 temporal operator $\hat{\tau}_2$ on the time. The time-dependent quasi momentum has $\hat{U} \mathbf{k}(t)=-\mathbf{k}(t)$. The interband current under $\hat{U}$ takes (see Appendix D for details)
\begin{equation}
\begin{aligned}
\hat{U} \mathbf{J}_{n m}^a(\mathbf{K}, t) \hat{U}^{\dagger}&=\mathbf{J}_{n m}^a\left(-\mathbf{K}, \hat{\tau}_2 t\right)\\
&=-\mathbf{J}_{n m}^a(\mathbf{K}, t) e^{2 i\left[S_{\mathrm{shift}}\left(\mathbf{K}, t, t^{\prime}\right)+S_{\Delta \mathrm{TDP}}(\mathbf{k}(t))\right]}.
\end{aligned}
\end{equation}
By comparing Eqs. (11) and (12), it can be found that the impact of $\hat{P}$ and  $\hat{U}$ on the photocurrent differs by the shift phase and $\Delta \mathrm{TDP}$. The shift phase vanish in crystals with both time-reversal symmetry and inversion symmetry (see Appendix D for derivations). However, by time-reversal symmetry alone, a phase mismatch between the dynamical phase and shift phase as well as between dynamical phase and $\Delta \mathrm{TDP}$ will be caused, and completely destructive interference cannot be formed. Since the inversion symmetry results in pure odd-order harmonic generation, the shift phase and $\Delta \mathrm{TDP}$ should be crucial factors for even-order harmonic generation in crystals with time-reversal symmetry.

A channel of the three-step model in strong-field physics includes excitation, acceleration and re-collision processes of an electron-hole pair. Let’s further consider two interference channels driven by monochromatic laser field as Fig. 1(c) shows. When inversion symmetry exists, the two channels are identical with an interval of half an optical cycle (black arrows indicate), their interference leads to pure odd-order harmonics. If the system only possesses time-reversal symmetry, then these two channels are going to be different (magenta arrows indicate). The band dispersion and transition dipole amplitude remain symmetric in $k$-space due to the protection of time-reversal symmetry. However, due to the existence of the shift vector in non-centrosymmetric crystals, electrons need to do extra work in the photoelectric field when they take interband transitions \cite{14}. Thus, the external light field equivalently modulate the energy curve of electrons like coulomb field, and forming the laser-dressed effective bands with symmetry breaking in $k$-space (magenta dashed curves in Fig. 1(c)). Therefore, extra shift phase accumulates in addition to the dynamical phase when pairs of dipoles perform intraband motions. The interference condition of pure odd-order harmonics is broken and even-order harmonics can be produced. The movement of electron-hole pairs on the effective bands simultaneously accumulates the dynamical phase and shift phase, which have fully equivalent effects on the photocurrent and together constitute its phase: 
\begin{equation}
S_{\mathrm{J}}\left(\mathbf{K}, t, t^{\prime}\right)=S_{\mathrm{dyn}}\left(\mathbf{K}, t, t^{\prime}\right)+S_{\mathrm{shift}}\left(\mathbf{K}, t, t^{\prime}\right)+S_{\Delta \mathrm{TDP}}(\mathbf{k}(t)),
\end{equation}

This fundamental image of interference involving two channels can be easily  generalized to multiple channels. Therefore, the coherence process of SSHG can be clearly described by coherent channels of dipoles on the laser-dressed effective bands.

\subsection{Using Circular Dichroism to Discriminate Time-Reversal Symmetry Breaking}

Based on above discussions, we continue to search for rules of SSHG that could be caused by time-reversal symmetry. The inversion symmetry breaking of crystals can be judged by even-order harmonic generation, which arises from interference of non-equivalent currents between two adjacent half cycles. Similarly, we can utilize laser ﬁelds with opposite helicities to ﬁnd evidence of time-reversal symmetry breaking. The helicity of elliptically polarized light can be ﬂipped by $\hat{T}$ without considering the Poynting vector of lights. It is found that the transport processes of charge carriers are diﬀerent in magnetic materials driven by lasers with diﬀerent helicity. In experiments, the magnetic circular dichroism of nonlinear optical response has been used to record the magnetic switching of materials \cite{36,37}. The circular dichroism caused by laser ﬁelds also have access to selective excitation of spin, valley and chirality of electron states \cite{38,39,40,41}. 

In the following, we will demonstrate that the circular dichroism of SSHG is directly related to the time-reversal symmetry breaking of crystal by our theoretical method. Under the time-reversal transformation, the initial right-hand helically polarized laser ($\sigma_{+}$) is changed to left-hand helically polarized laser ($\sigma_{-}$), which have $\mathbf{E}_{\sigma_{+}}(t)=\mathbf{E}_{\sigma_{-}}(-t)$, and $\mathbf{A}_{\sigma_{+}}(t)=-\mathbf{A}_{\sigma_{-}}(-t)$. Each component of the photocurrent phase has the following transformation relation,
\begin{equation}
\begin{aligned}
\hat{T} S_{\mathrm{dyn}}\left(\mathbf{K}, \mathbf{A}_{\sigma_{+}}, t, t^{\prime}\right) \hat{T}^{\dagger} & =S_{\mathrm{dyn}}\left(-\mathbf{K}, \mathbf{A}_{\sigma_{-}},-t,-t^{\prime}\right)\\
&=-S_{\mathrm{dyn}}\left(\mathbf{K}, \mathbf{A}_{\sigma_{+}}, t, t^{\prime}\right), 
\end{aligned}
\end{equation}
\begin{equation}
\begin{aligned}
\hat{T} S_{\mathrm{shift}}\left(\mathbf{K}, \mathbf{A}_{\sigma_{+}}, t, t^{\prime}\right) \hat{T}^{\dagger} & =S_{\mathrm{shift}}\left(-\mathbf{K}, \mathbf{A}_{\sigma_{-}},-t,-t^{\prime}\right)\\
&=-S_{\mathrm{shift}}\left(\mathbf{K}, \mathbf{A}_{\sigma_{+}}, t, t^{\prime}\right), 
\end{aligned}
\end{equation}
\begin{equation}
\begin{aligned}
\hat{T} S_{\Delta \mathrm{TDP}}\left(\mathbf{K}+\mathbf{A}_{\sigma_{+}}(t)\right) \hat{T}^{\dagger} & =S_{\Delta \mathrm{TDP}}\left(-\mathbf{K}+\mathbf{A}_{\sigma_{-}}(-t)\right)\\
&=-S_{\Delta \mathrm{TDP}}\left(\mathbf{K}+\mathbf{A}_{\sigma_{+}}(t)\right).
\end{aligned}
\end{equation}
The dynamical phase, shift phase and $\Delta \mathrm{TDP}$ all reverse signs under $\hat{T}$, but this cannot be achieved by any pure point-group symmetry. Accordingly, the interband current generated by helically polarized lasers transforms as (see Appendix D for derivations)
\begin{equation}
\begin{aligned}
\hat{T} \mathbf{J}_{n m, \sigma_{+}}^a(\mathbf{K}, t) \hat{T}^{\dagger}&=\mathbf{J}_{n m, \sigma_{-}}^a(-\mathbf{K},-t)\\
&=-\mathbf{J}_{n m, \sigma_{+}}^{a, *}(\mathbf{K}, t) \frac{f_{n m}(-\mathbf{K},-t)}{f_{n m}(\mathbf{K}, t)} .
\end{aligned}
\end{equation}

If we assume that the difference of Fermi-Dirac distribution is time-reversal invariant (i.e., $f_{n m}(-\mathbf{K},-t)=f_{n m}(\mathbf{K}, t)$), the interband current has $\hat{T} \mathbf{J}_{n m, \sigma_{+}}^a(t) \hat{T}^{\dagger}=$ $-\mathbf{J}_{n m, \sigma_{+}}^{a, *}(t)$. This assumption can work in the perturbation region with low-order changing rate of electron distribution. For the intraband current, the same conclusion can be derived. Then we have $\hat{T} \mathbf{J}_{\sigma_{+}}^a(t) \hat{T}^{\dagger}=\mathbf{J}_{\sigma_{-}}^a(-t)=-\mathbf{J}_{\sigma_{+}}^a(t)$, so lasers with opposite helicity can produce SSHG with the same intensity. We deduce that, under the protection of the time-reversal symmetry of crystals, the intensity of low-order nonlinear optical response does not show the circular dichroism. In contrast, the circular dichroism emerges when magnetic materials are considered. It allows us to use helically polarized laser fields to detect the magnetism of crystals.

\section{Model Calculation}

We have theoretically revealed the role of the  $\Delta \mathrm{TDP}$ and the shift phase in symmetric rules of SSHG. Now we perform numerical calculations based on tight-binding models including the graphene, h-BN and Haldane model \cite{42} to justify our theoretical insight.

\subsection{Orientation Dependence and Polarization Characteristics}

Our discussion focuses on tight-binding models with honeycomb lattice due to its universality. Considering the hopping to the nearest-neighbor sites, the Hamiltonian is

\begin{equation}
H=t_1 \sum_{\langle i, j\rangle} c_i^{\dagger} c_j,
\end{equation}
where $i$, $j$ denote different sublattices. This is the simplest two-band Hamiltonian used to describe the graphene, which is subject to $D_{6 h}$ point-group and time-reversal symmetries. For calculations, we set the lattice constant to 2.5 $\mathrm{\AA}$ and the nearestneighbor hopping $t_1$ to 2.33 $\mathrm{eV}$. The peak intensity of the driving laser we selected is $1.2 \times 10^{12} \mathrm{~W} / \mathrm{cm}^2$, wavelength is 1.9 $\mu \mathrm{m}$, and full width at half maximum is 55 $\mathrm{fs}$ in a Gaussian envelope.

Figure 2 shows the polarization characteristics of harmonics parallel and perpendicular to linearly polarized laser field as a function of crystal orientation. The orientation angle $\theta$ is set to 0° when the laser is along $\Gamma-K$ direction. Pure odd-order harmonics are generated due to the inversion symmetry of graphene [see Figs. 2(a) and 2(b)]. The $C_{6}$ axis out of plane leads to the orientation periodicity of 60° for all harmonics.

\begin{figure}

    \includegraphics[width=3.4in,angle=0]{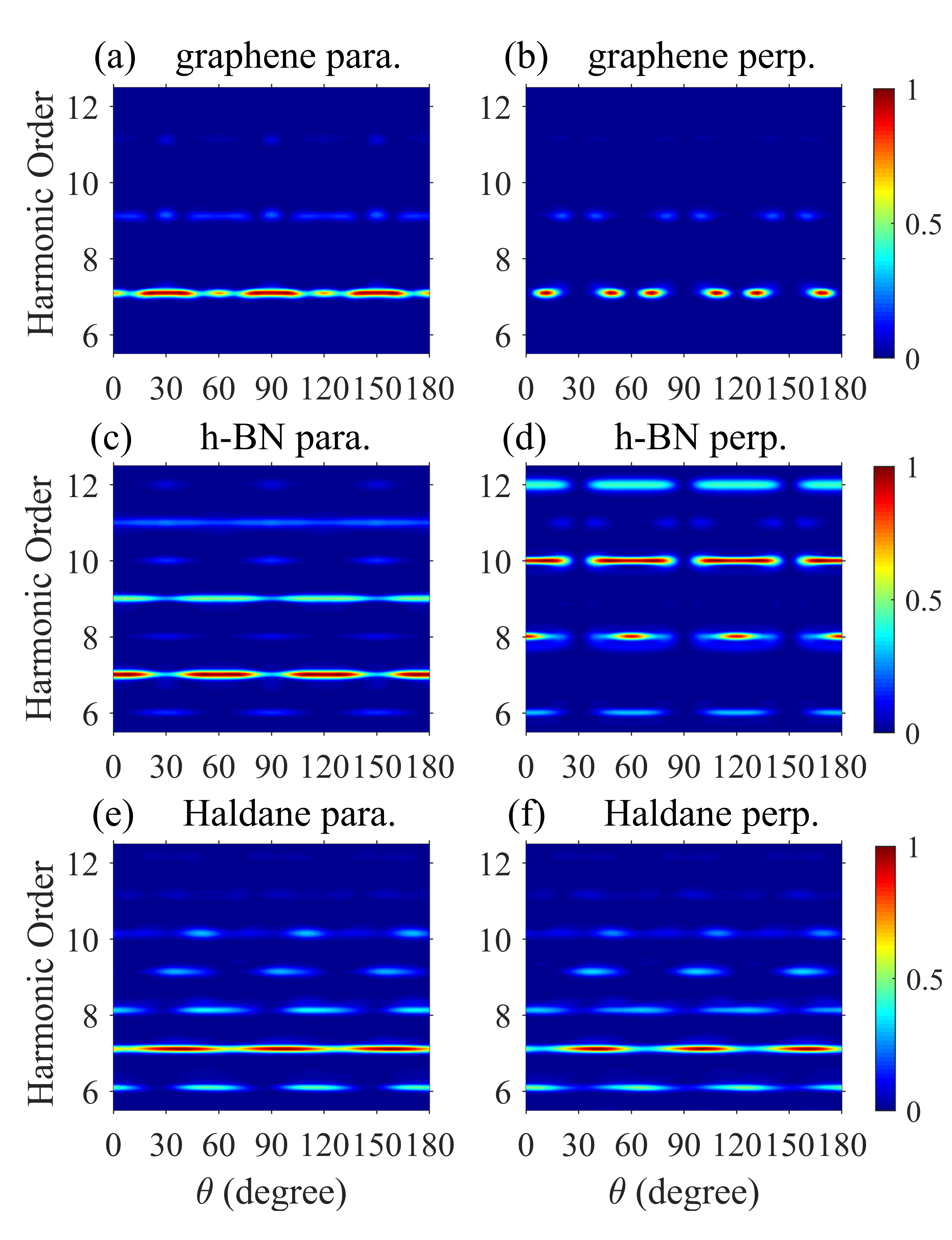}
    \caption{ Orientation dependence and polarization characteristics. Parallel and perpendicular components of the SSHG for (a,b) graphene, (c,d) h-BN, and (e,f) Haldane model. 
}\label{Fig_2}
\end{figure}

Let’s break the inversion symmetry by introducing different on-site energy for 
adjacent atoms, the Hamiltonian is
\begin{equation}
H=t_1 \sum_{\langle i, j\rangle} c_i^{\dagger} c_j+M_0 \sum_i \epsilon_i c_i^{\dagger} c_i,
\end{equation}
in which $\epsilon_i$= ±1 for different atoms. It can be used to describe hexagonal boron nitride (h-BN), its point-group symmetry is reduced to $D_{3 h}$. Different from the 
graphene, even harmonics are generated due to the inversion symmetry breaking (see Figs. 2(c) and 2(d)). The inversion symmetry breaking term $M_0$ is set to 1.96 eV. The 
interference processes in time domain can be reflected by time-frequency analysis spectra. Combining Figs. 3(b) and 3(d), we now know that even-order harmonics result from the difference of harmonic radiations between adjacent half cycles. By analyzing 
Eq. (13), the inverted laser field every half optical cycle cannot change the dynamical phase. However, the shift phase as well as $\Delta \mathrm{TDP}$ formed by the two unequal 
interference channels are different, which are responsible for the different temporal harmonic radiations and even-order harmonic generation in non-centrosymmetric crystals (see channel 1 and channel 2 in Figs. 3(c) and 3(e), which are same for graphene but different for h-BN). In Fig. 3, only parallel component of photocurents is considered, thus the $\Delta \mathrm{TDP}$ is vanish here but exist for other polarization directions.

\begin{figure}[h]
    \includegraphics[width=3.4in,angle=0]{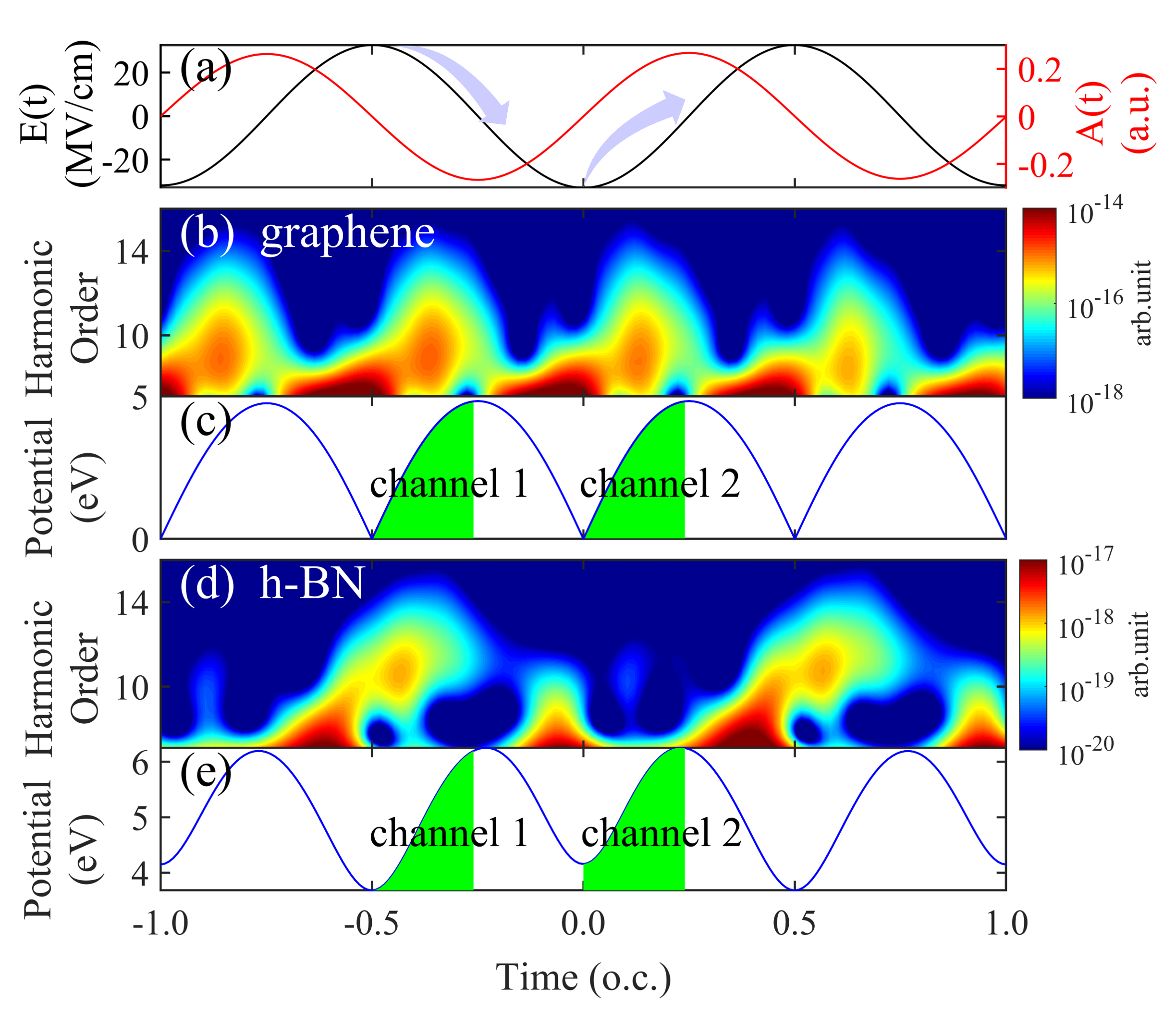}
    \caption{Effect of photocurrent phase on harmonic radiation in time domain. (a) Electric field and 
vector potential of laser, the arrow indicates two adjacent electron channels, separated by half optical 
cycle. (b,d) Time-frequency analysis for graphene and h-BN. (c,e) Time-dependent laser-dressed 
effective potential between electron and hole excited at $K$-point for graphene and h-BN. Integrals 
of the effective potentials in the time domain are the photocurrent phases, which are marked by 
green regions for two interference channels. Laser polarization is along $\Gamma-M$ direction, the polarization 
of harmonics is parallel to the laser.}\label{Fig_3}
\end{figure}

In addition, the mirror symmetry can also induce destructive interference of harmonics as we derived by Eq. (10). When the driving laser is oriented parallel to the mirror plane $\left(\theta=30^{\circ} \pm 60^{\circ} l, l \in \mathbb{N}\right)$, there are no harmonics of perpendicular polarization because the currents cancel each other out. When the laser field is perpendicular to the mirror plane $\left(\theta=60^{\circ} \pm 60^{\circ} l, l \in \mathbb{N}\right)$, it strictly follows that only odd-order harmonics are generated for the parallel polarization and even-order harmonics for the perpendicular polarization. In addition, notice that although the h-BN only has an in-plane $C_3$ symmetry, the harmonics can form 6-fold orientation periodicity. This can be completely attributed to the multi-cycle driving field usually used. The reversal of a multi-cycle laser field, equivalent to its carrier envelope phase shifts $\pi$, does not affect the overall intensity of harmonics. If a few-cycle driving field is used, such extra 2-fold orientation periodicity of harmonics cannot be observed. Therefore, a few-cycle driving probe is needed to determine the rotation axis of crystals.

 To further discuss the effect of the mirror symmetry and time-reversal symmetry breaking, we consider the Haldane model with magnetic phase on the next-nearest neighbor sites. The Hamiltonian is expanded as
 \begin{equation}
H=t_1 \sum_{\langle i, j\rangle} c_i^{\dagger} c_j+M \sum_i \epsilon_i c_i^{\dagger} c_i+t_2 \sum_{\langle\langle i, j\rangle\rangle} e^{-i v_{i j} \varphi} c_i^{\dagger} c_j,
\end{equation}
where $v_{i j}$=±1 depending on the kind of atoms that the hopping takes place between. Here, we do not care about the optical behavior of its topological properties, the complex hopping strength is considered only to break the mirror and time-reversal symmetries. The basic point-group symmetry of this Hamiltonian is reduced to $C_{3 h}$. The complex hopping strength $t_2$=0.63 eV, its phase $\varphi=\frac{\pi}{4}$. Since the mirrors are broken, the stable destructive and constructive interference that occur in the h-BN case vanishes in the Haldane model (see $\theta=\pm30^{\circ} l, l \in \mathbb{N}$ in Figs. 2(e) and 2(f)). The spectra keep the orientation periodicity of 60°, which arises from the $C_{3}$ axis and the multi-cycle driving field. 

 \subsection{Ellipticity Dependence}

The helicity dependence of SSHG can be used to probe molecular chirality, which is attributed to the circular dichroism of chiral molecules \cite{38}. The similar thing happens in crystals, the mirror symmetry can protect the harmonics from the circular dichroism. Let’s apply a helically polarized laser to a crystal with mirror symmetry. Under the mirror reflection, the applied right-hand helically polarized ($\sigma_{+}$) laser is changed to left-hand helically polarized ($\sigma_{-}$) laser. The photocurrent has
 \begin{equation}
\hat{M}_c \mathbf{J}_{\sigma_{+}}^a(t) \hat{M}_c^{\dagger}=\mathbf{J}_{\sigma_{-}}^{M_c^{-1} a}(t)=M_{c a} \mathbf{J}_{\sigma_{+}}^a(t),
 \end{equation}
where $c$ is the normal direction of mirror plane, and $M_{c a}=e^{i\left(\phi_{n m}^{M_c^{-1} a}-\phi_{n m}^a\right)}$. In other words, the mirror reflection of the laser-crystal system directly causes the reflection of photocurrent. Therefore, there is no circular dichroism in SSHG from crystals with mirror symmetry. As Fig. 4(a) shows, the harmonic spectrum of h-BN driven by circularly polarized lights with inverse helicities have the same intensity. Due to the $C_3$ symmetry of the h-BN and the circularly polarized driving laser we used here, only $3 l \pm 1(l \in \mathbb{N})$ harmonic orders are allowed.
 \begin{figure}[h]
    \includegraphics[width=3.4in,angle=0]{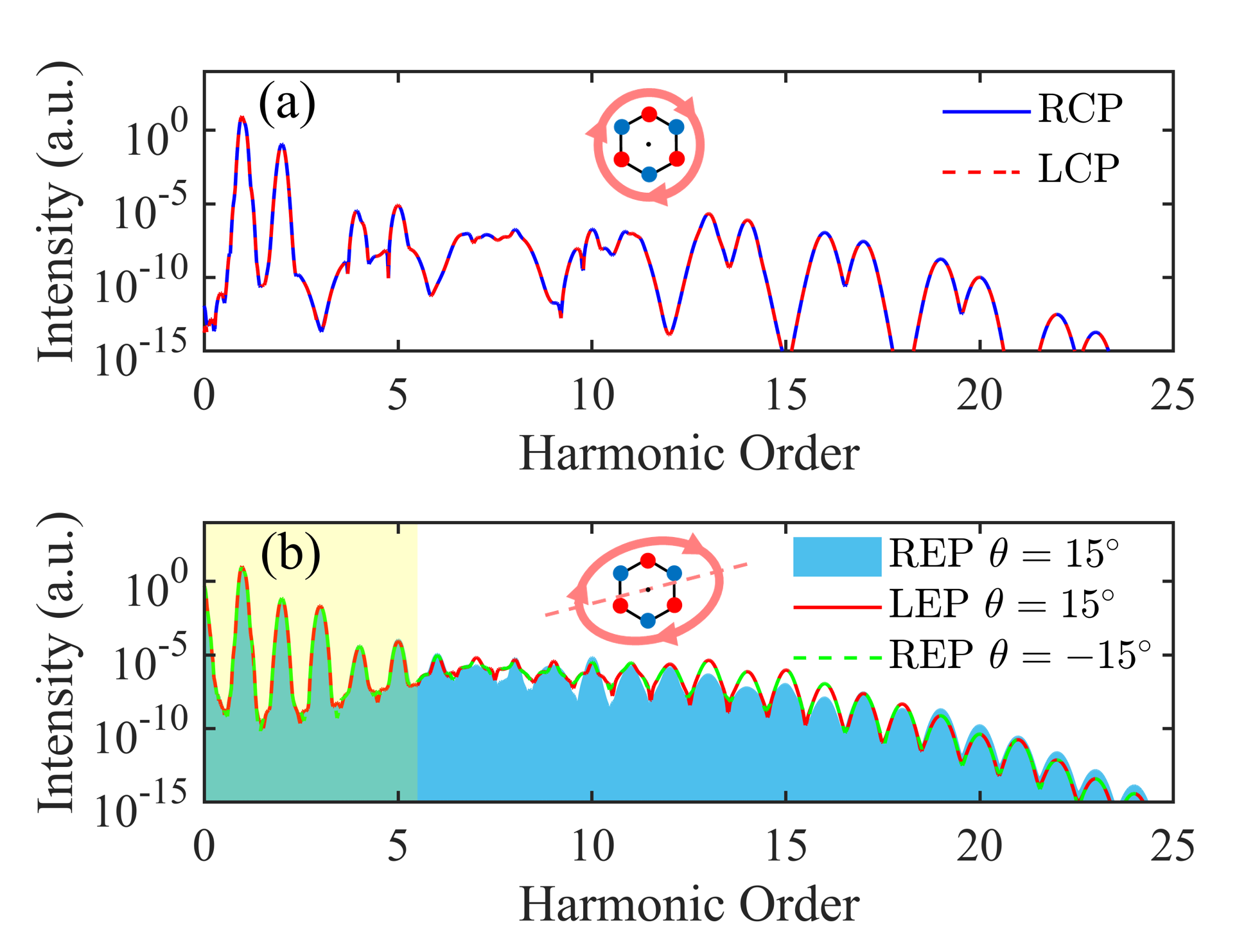}
    \caption{(a) SSHG of h-BN driven by right-handed circularly polarized (RCP) and left-handed circularly polarized (LCP) light. (b) SSHG of h-BN driven by elliptically polarized light (the ellipticity of REP is 0.5, and LEP is -0.5), $\theta$ is the orientation angle between the main axis of ellipse and $\Gamma-K$ direction. The harmonic intensity driven by left-handed elliptically polarized (LEP) light with $\theta$=15° (red solid line) is the same as that driven by right-handed elliptically polarized (REP) light with $\theta$=$-$15° (green dashed line), but the REP case with $\theta$=15° (blue area) is offset from them in the high-order region. }\label{Fig_4}
\end{figure}
However, an asymmetric profile of elliptically dependent SSHG from cubic  crystals has been demonstrated in Refs. \cite{43,44,45}, which was explained by the coupled  intraband and interband dynamics. Here, we attribute this asymmetry to the mirror 
reflection mismatch between light field and crystal. The circular dichroism may be  induced if only the helicity of elliptically polarized light is reversed while its orientation 
angle remains (see the difference in high-order region between the blue area and red solid line in Fig. 4(b)). When we reflect the laser-crystal system simultaneously, the  circular dichroism will vanish (see the red solid line and green dashed line in Fig. 4(b)).

The time-reversal symmetry can also reverse the helicity of in-plane laser fields. We can use the circular dichroism of harmonics to identify the time-reversal symmetry of crystals. We note the area shaded yellow in Fig. 4(b) that the low-order harmonics, nearly below bandgap, never exhibit circular dichroism in h-BN. However, a completely different phenomenon emerges in the Haldane model with breaking time-reversal symmetry. 

\begin{figure}
    \includegraphics[width=3.4in,angle=0]{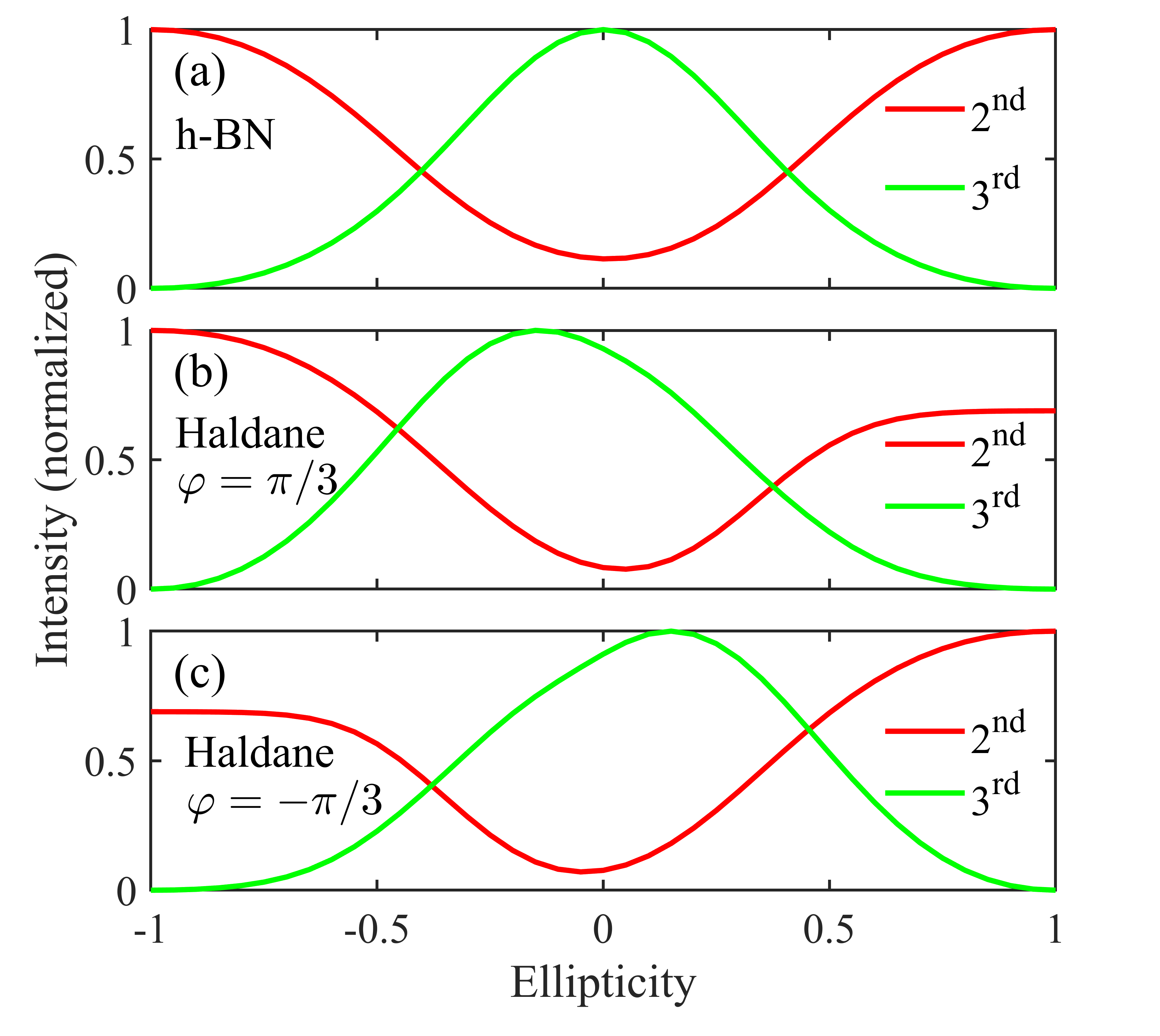}
    \caption{Ellipticity dependence of low-order harmonic intensity for h-BN (a) and Haldane model with $\varphi={\pi}/{3}$ (b), $\varphi=-{\pi}/{3}$ (c). The orientation angle $\theta$ is set to be 15°, other parameters of the models and laser are the same as Fig. 2. The intensity values are normalized for the 2$^{nd}$ and 3$^{rd}$ harmonics. }\label{Fig_4}
\end{figure}

We deduce from Eq. (17) that, the time-reversal symmetry of crystals prevents the low-order harmonics from circular dichroism. In order to exclude the effect of mirror symmetry, we use an elliptically polarized laser and set the orientation angle to 15°. Expectedly, the low-order harmonics of h-BN keep perfect ellipticity-dependent symmetry [see Fig. 5(a)]. However, due to the absence of the time-reversal symmetry, asymmetric ellipticity dependence can be clearly seen in the Haldane model [see Fig. 5(b)]. Thus, the circular dichroism of harmonics can be used to identify magnetic materials. 

Then we consider the time-reversal enantiomer by flipping the magnetic flux of the Haldane model. As Figs. 5(b) and 5(c) show, when the phase of the flux is inverted, the ellipticity dependence is reversed exactly as well. This is equivalent to performing $\hat{T}$ on the laser-crystal system, but the intensity of low-order harmonics is unaltered. Combined with ultrafast time-resolved spectra, it is promising that the low-order harmonics can be used to observe the magnetization degree of materials or ultrafast spin dynamics \cite{46,47}. 

\section{Conclusions}
In summary, we reorganize the expression of photocurrent under the dipole approximation and reveal the indispensable roles of the shift phase and $\Delta \mathrm{TDP}$. A more systematic picture of the SSHG in external laser fields has been obtained. Firstly, we point out that the selection rule of SSHG is determined by the phase difference of transition dipole moments under point-group symmetry operation. Secondly, similar to the dynamical phase caused by the Coulomb field, the shift phase is induced by the instantaneous potential of the oscillating laser, the motion of electrons on the laser-dressed effective band deserves to be a complementary strong-field physical image to study the SSHG. For examples, when we reconstruct the band structure of non-centrosymmetric crystals and consider the propagation effect of SHHG or its phase matching condition, the effects of the shift phase and $\Delta \mathrm{TDP}$ are non-negligible \cite{4,26}. Our framework can help us to understand the microscopic mechanism of the selection rules, orientation dependence, polarization characteristics, time-frequency analysis and ellipticity dependence of SSHG. Our two-channel image can also be used to explain the cause of the valley polarization and valley Hall effect \cite{38}. Since the shift phase and the berry phase have a great relationship, we expect that this framework can promote SSHG for the characterization of topological band geometry \cite{14,50}. Thirdly, we also point out that the low-order harmonics do not have circular dichroism in nonmagnetic systems, which is important for identifying the magnetization degree of materials based on strong-field nonlinear optics. Strong-field ultrafast nonlinear effect is a vital tool for exploring crystal structure and electron spin dynamics \cite{37,48,49}. Our theory and conclusions can be extended to discuss optical phenomena about the magnetic point groups. In addition, our discussion is limited to the photocurrent under the electric dipole approximation, which can be generalized to the electric quadrupole and magnetic dipole regions. Last but not the least, the single-electron framework is no longer suitable for systems with strongly correlated interaction, thus establishing efficient model involving quasiparticle transitions becomes particularly urgent.

\begin{acknowledgments}

We gratefully acknowledge the support of the National Key Research and Development Program of China (2022YFA1604301, 2018YFA0305700), the NSF of China (11974185, 11704187, 11834004, 11925408, 11921004, 12188101, 12174195, 12304378), the Natural Science Foundation of Jiangsu Province (Grant No. BK20170032), Fundamental Research Funds for the Central Universities (No. 30920021153), the Project funded by China Postdoctoral Science Foundation (Grant No. 2019M661841), the Postgraduate Research Practice Innovation Program of Jiangsu Province (KYCX22$\_$0419), the Strategic Priority Research Program of CAS (XDB33000000), and the K. C. Wong Education Foundation (Grant No. GJTD-2018-01).The authors thank Guanglu Yuan, Hanqi Pi, Xiangyu Zhang, Zishao Wang and Lihan Chi for useful discussions.
\end{acknowledgments}

%The authors declare no competing financial interests. 
\section*{APPENDIX A: Derivation of Photocurrent}
\setcounter{equation}{0} 
\renewcommand{\theequation}{A\arabic{equation}}
Based on the single-electron and dipole approximations, we can directly write the time-dependent Hamiltonian of matter interacting with an external laser field (atomic units are used throughout unless otherwise stated), 
\begin{equation}
\begin{aligned}
		&\hat{H}\left(t\right)={\hat{H}}_0\left(t\right)+{\hat{H}}_l\left(t\right),\\
		&{\hat{H}}_0=-\frac{1}{2}\mathrm{\nabla}_\mathbf{\mathbf{r}}^2+\hat{V}\left(\mathbf{r}\right),\\
&{\hat{H}}_l\left(t\right)=\hat{\mathbf{r}}\cdot\mathbf{E}\left(t\right),
\end{aligned}
\end{equation}
where $\nabla_{\mathbf{r}}^2$ is the Laplace operator with respect to the electronic coordinate operator $\hat{\mathbf{r}}$, $V(\mathbf{r})$ is the Coulomb potential, $\mathbf{E}(t)$ is the applied electric field. It can be seen that the laser field only acts on the coordinate operators of a single electron without changing the Hamiltonian $\hat{H}_0$ of the initial system. In crystals, $V(\mathbf{r})$ has the translational symmetry, and electrons in the periodic lattice potential can be described by a wave packet composed of Bloch waves,
\begin{equation}
\psi(\mathbf{r}, t)=\frac{1}{N_{\mathrm{c}}^{1 / 2}} \sum_m \int_{\mathrm{BZ}} d \mathbf{k} a_{m, \mathbf{k}}(t) \phi_{m, \mathbf{k}}(\mathbf{r})
\end{equation}
with $\phi_{m, \mathbf{k}}(\mathbf{r})=e^{i \mathbf{k r}} u_{m, \mathbf{k}}(\mathbf{r})$, here $u_{m, \mathbf{k}}(\mathbf{r})$ is the periodic part of the Bloch wave function. $N_{\mathrm{c}}$ is the total number of unit cells, $a_{m, \mathbf{k}}(t)$ is the time-dependent probability amplitude of the Bloch wave. By using the time-dependent Schrödinger equation $i \frac{\partial \psi(\mathbf{r}, t)}{\partial t}=H(t) \psi(\mathbf{r}, t)$, we can obtain
\begin{equation}
\begin{aligned}
i \frac{\partial a_{n, \mathbf{k}^{\prime}}(t)}{\partial t}=\ &\varepsilon_{n, \mathbf{k}^{\prime}} a_{n, \mathbf{k}^{\prime}}(t) \\
&+\mathbf{E}(t) \cdot \sum_m \int_{\mathrm{BZ}} d \mathbf{k} a_{m, \mathbf{k}}(t)\left\langle\phi_{n, \mathbf{k}^{\prime}}|\hat{\mathbf{r}}| \phi_{m, \mathbf{k}}\right\rangle,
\end{aligned}
\end{equation}
where the position operator under Bloch basis can be rewritten as $\left\langle\phi_{n, \mathbf{k}^{\prime}}|\hat{\mathbf{r}}| \phi_{m, \mathbf{k}}\right\rangle=$ $\delta\left(\mathbf{k}^{\prime}-\mathbf{k}\right)\left(-i \delta_{n, m} \partial_{\mathbf{k}^{\prime}}+\mathbf{d}_{n m}(\mathbf{k})\right)$, here the transition dipole matrix element $\mathbf{d}_{n m}(\mathbf{k})=i\left\langle u_{n, \mathbf{k}}\left|\partial_{\mathbf{k}}\right| u_{m, \mathbf{k}}\right\rangle$ is introduced to describe the polarization of electron-hole pairs. Eq. (A3) can be converted to the Houston basis after gauge transformations,
\begin{equation}
\begin{aligned}
i \frac{\partial a_{n, \mathbf{k}(t)}(t)}{\partial t}=\ &\mathbf{E}(t) \cdot \sum_m \mathbf{d}_{n m}(\mathbf{k}(t)) a_{m, \mathbf{k}(t)}(t) \\
&\times e^{i \int_{-\infty}^t \varepsilon_{n m}(\mathbf{k}(\tau)) d \tau},
\end{aligned}
\end{equation}
where $\varepsilon_{n m}(\mathbf{k})=\varepsilon_n(\mathbf{k})-\varepsilon_m(\mathbf{k})$ is the energy difference between bands. The quasi momentum $\mathbf{k}(t)$ of electrons changes adiabatically with the laser field, and the evolution relationship is $\mathbf{k}(t)=\mathbf{K}+\mathbf{A}(t)$. $\mathbf{K}$ is the canonical momentum of the crystal in the absence of field. This is a multi-band coupling equation, in which the probability amplitude of the electron in the $n$th eigenstate is related to other states through the transition dipole moments. Eq. (A4) can be regarded as a linear superposition of these transition processes, which form a statistical ensemble.

We introduce density matrix $\rho_{n m}(\mathbf{K}, t)=a_{n, \mathbf{k}(t)}^{\dagger}(t) a_{m, \mathbf{k}(t)}(t)$ to describe the time-dependent evolution of the electronic population. For simplicity, here we consider the population transition that occur mainly between two bands, which can always be described by a two-band equation. The densities of interband and intraband currents obey (we ignore the dephasing time related to coupling between particles):
\begin{widetext}
\begin{equation}
\begin{aligned}
{\dot{\rho}}_{nm}\left(\mathbf{K},t\right)=&-i\mathbf{E}\left(t\right)\cdot\left[\left(\mathbf{d}_{mm}\left(\mathbf{k}\left(t\right)\right)-\mathbf{d}_{nn}\left(\mathbf{k}\left(t\right)\right)\right)\rho_{nm}\left(\mathbf{K},t\right)+\mathbf{d}_{mn}\left(\mathbf{k}\left(t\right)\right)f_{nm}\left(\mathbf{K},t\right)e^{i\int_{-\infty}^{t}{\varepsilon_{mn}\left(\mathbf{k}\left(\tau\right)\right)}d\tau}\right],
\end{aligned}
\end{equation}
\begin{equation}
\begin{aligned}
\dot{\rho}_{n n}(\mathbf{K}, t)= & -i \mathbf{E}(t) \cdot \mathbf{d}_{n m}(\mathbf{k}(t)) \rho_{n m}(\mathbf{K}, t) e^{i \int_{-\infty}^t \varepsilon_{n m}(\mathbf{k}(\tau)) d \tau}+\mathrm { c. c. },
\end{aligned}
\end{equation}
\end{widetext}
where $f_{n m}(\mathbf{K}, t)=\rho_{n n}(\mathbf{K}, t)-\rho_{m m}(\mathbf{K}, t)$ is the difference of Fermi-Dirac distribution, and the band index $n \neq m$. Equations (S7) and (S8) are results of the coupling between the adiabatic evolution and non-adiabatic tunneling process of the electron density. The time-dependent photocurrent can be divided into interband and intraband components.
\begin{equation}
\begin{aligned}
& \mathbf{J}_{n m}(t)=-\frac{1}{N_{\mathrm{c}}} \sum_{\mathbf{K} \in \mathrm{BZ}} \rho_{n m}(\mathbf{K}, t) \mathbf{p}_{n m}(\mathbf{k}(t)), 
\end{aligned}
\end{equation}
\begin{equation}
\begin{aligned}
& \mathbf{J}_{n n}(t)=-\frac{1}{N_{\mathrm{c}}} \sum_{\mathbf{K} \in \mathrm{BZ}} \rho_{n n}(\mathbf{K}, t) \mathbf{p}_{n n}(\mathbf{k}(t)),
\end{aligned}
\end{equation}
where the momentum operator can be given by $\hat{\mathbf{p}}(\mathbf{k}, t)=\partial_{\mathbf{k}} \hat{H}(\mathbf{k}, t)$. In this paper, we assume the evolution of the electron population has no effect on the Coulomb potential. Therefore, the original Hilbert space does not change with the addition of laser fields. The momentum matrix element takes the form $\mathbf{p}_{n m}(\mathbf{k})=\left\langle u_{n, \mathbf{k}}|\hat{\mathbf{p}}| u_{m, \mathbf{k}}\right\rangle$, which can be calculated by
\begin{equation}
\begin{aligned}
\mathbf{p}_{n m}(\mathbf{k}) & =i \varepsilon_{n m}(\mathbf{k}) \mathbf{d}_{n m}(\mathbf{k}), n \neq m, 
\end{aligned}
\end{equation}
\begin{equation}
\begin{aligned}
\mathbf{p}_{n n}(\mathbf{k}) & =\partial_{\mathbf{k}} \varepsilon_n(\mathbf{k}) .
\end{aligned}
\end{equation}
The anomalous velocity induced by the Berry curvature has been included in Eq. (A7). Then, we can obtain expressions for the interband and intraband currents as Eqs. (1) and (2).

\section*{APPENDIX B: Transformation of Transition Dipole under Point-Group Symmetry}
\setcounter{equation}{0} 
\renewcommand{\theequation}{B\arabic{equation}}
Consider a point-group symmetry operation $\hat{G}$ on the Bloch state of electrons,
\begin{equation}
\hat{G}\left|u_{n, \mathbf{k}}\right\rangle=\left|u_{n, G^{-1} \mathbf{k}}\right\rangle \mathrm {. }
\end{equation}
For the transition dipole matrix element, we have
\begin{equation}
\begin{aligned}
\hat{G} \mathbf{d}_{n m}^a(\mathbf{k})\hat{G}^{\dagger} & =i \hat{G}\left\langle u_{n, \mathbf{k}}\left|\partial_{\mathbf{k}_a}\right| u_{m, \mathbf{k}}\right\rangle \hat{G}^{\dagger} \\
& =i\left\langle u_{n, G^{-1} \mathbf{k}}\left|\hat{G} \partial_{\mathbf{k}_a} \hat{G}^{\dagger}\right| u_{m, G^{-1} \mathbf{k}}\right\rangle \\
& =i\left\langle u_{n, G^{-1} \mathbf{k}}\left|\partial_{\left(G^{-1} \mathbf{k}\right)_{a^{\prime}}}\right| u_{m, G^{-1} \mathbf{k}}\right\rangle \\
& =\mathbf{d}_{n m}^{a^{\prime}}\left(G^{-1} \mathbf{k}\right),
\end{aligned}
\end{equation}
where $a, a^{\prime}=G^{-1} a$ denote the directions of the transition dipole moments. If the crystal has $G$ symmetry, its Hamiltonian satisfies $\hat{G} \hat{H}=\hat{H} \hat{G}$, the Bloch wave is thus the eigenstate of $\hat{G}$ as well. Since $\hat{G}$ is unitary, its eigenvalues are complex numbers of modulo 1 :
\begin{equation}
\hat{G}\left|u_{n, \mathbf{k}}\right\rangle=e^{i \phi_G}\left|u_{n, \mathbf{k}}\right\rangle .
\end{equation}
Thus, we obtain $\left|u_{n, G^{-1} \mathbf{k}}\right\rangle=e^{i \phi_G}\left|u_{n, \mathbf{k}}\right\rangle$. Then,
\begin{equation}
\begin{aligned}
\mathbf{d}_{n m}^{a^{\prime}}\left(G^{-1} \mathbf{k}\right) & =i\left\langle u_{n, \mathbf{k}}\left|\partial_{\left(G^{-1} \mathbf{k}\right)_{a^{\prime}}}\right| u_{m, \mathbf{k}}\right\rangle \\
& =i e^{i\left[\phi_{n m}^{a^{\prime}}(\mathbf{k}(t))-\phi_{n m}^a(\mathbf{k}(t))\right]} \left\langle u_{n, \mathbf{k}}\left|\partial_{\mathbf{k}_a}\right| u_{m, \mathbf{k}}\right\rangle   \\
& =G_{a^{\prime} a} \mathbf{d}_{n m}^a(\mathbf{k}),
\end{aligned}
\end{equation}
in which $G_{a^{\prime} a}$ is just the $\Delta$TDP between $a$ and $a^{\prime}$ direction under $\hat{G}$. Moreover, scalar quantities such as the dynamical phase and shift phase are invariant under $\hat{G}$.

\section*{APPENDIX C: Typical Cases of Selection Rules of SHG}
\setcounter{equation}{0} 
\renewcommand{\theequation}{C\arabic{equation}}
\subsection*{C1. Mirror Symmetry}

\noindent A two-fold mirror dynamical symmetry can be obtained by exciting the dipole pairs with oscillating electric field which is symmetric about a mirror plane. Adjoining order-2 temporal operator $\hat{\tau}_2$ with the mirror reflection $\hat{M}_c$, we define the dynamical symmetry operation $\hat{F}=$ $\hat{M}_c \cdot \hat{\tau}_2$, where $c$ is the normal direction of the mirror plane. Monochromatic lights reverse along $c$ under $\hat{\tau}_2$ (i.e., $\hat{\tau}_2 \mathbf{E}^c(t)=-\mathbf{E}^c(t)$ ), so we have $\hat{F} \mathbf{k}(t)=M_c^{-1} \mathbf{K}+$ $\mathbf{A}\left(\hat{\tau}_2 t\right)=M_c^{-1} \mathbf{k}(t)$. According to Eq. (10), we derive that the interband photocurrent transforms as
\begin{equation}
\begin{aligned}
&\hat{F} \mathbf{J}_{n m}^a(\mathbf{K}, t) \hat{F}^{\dagger} \\
&=\mathbf{J}_{n m}^{M_c^{-1} a}\left(M_c^{-1} \mathbf{K}, \hat{\tau}_2 t\right) \\
&=-i \varepsilon_{n m}\left(M_c^{-1} \mathbf{k}(t)\right)\mathbf{d}_{n m}^{M_c^{-1} a}\left(M_c^{-1} \mathbf{k}(t)\right) \rho_{n m}\left(M_c^{-1} \mathbf{K}, \hat{\tau}_2 t\right) \\
&=-i \varepsilon_{n m}(\mathbf{k}(t))\mathbf{d}_{n m}^a(\mathbf{k}(t)) e^{i\left[\phi_{n m}^{M_c^{-1} a}(\mathbf{k}(t))-\phi_{n m}^a(\mathbf{k}(t))\right]} \\
&\ \ \ \  \times\rho_{n m}(\mathbf{K}, t).
\end{aligned}
\end{equation}

When $a \| c$, the $\Delta \mathrm{TDP}$ is $\phi_{n m}^{M_c^{-1} a}(\mathbf{k})-\phi_{n m}^a(\mathbf{k})=\pi$. Thus, the photocurrent satisfies $\hat{F} \mathbf{J}_{n m}^a(t) \hat{F}^{\dagger}=-\mathbf{J}_{n m}^a(t)$. Such reversal of the current only comes from the reflected dipole moment $\mathbf{d}_{n m}^a$. This Floquet system can be reduced to two pairs of electric dipoles with opposing polarization directions and separated by half an optical cycle (o.c.) in the time domain [see Fig. 1(a)].

The photons with frequency $\omega$ become interference enhanced if the current satisfies $\hat{F} \mathbf{J}_{n m}^a(\omega) \hat{F}^{\dagger}=\mathbf{J}_{n m}^a(\omega)$. Using the Fourier transform that $\mathbf{J}_{n m}^a(\omega)=$ $\int_{-\infty}^{+\infty} d t \mathbf{J}_{n m}^a(t) e^{i \omega t}$, we have
\begin{equation}
e^{i \omega \frac{T_0}{2}}=e^{i \pi} \Rightarrow \omega=(2 l+1) \omega_0, l \in \mathbb{N}
\end{equation}
where $\mathbb{N}$ denotes natural numbers. Considering similar transformation for Eq. (2) corresponding to intraband current, we can easily reach the same conclusion. Therefore, when the laser field is perpendicular to the mirror plane, only odd-order harmonics emit in the direction parallel to the driving field.

When $a \perp c$, the $\Delta \mathrm{TDP}$ is $\phi_{n m}^{M_c^{-1} a}(\mathbf{k})-\phi_{n m}^a(\mathbf{k})=0$. The photocurrent satisfies $\mathbf{J}^{M_c^{-1} a}\left(\hat{\tau}_2 t\right)=\mathbf{J}^a(t)$. This derives only even-order harmonic generation in the direction perpendicular to the driving field $\left(\omega=2 l \omega_0, l \in \mathbb{N}\right)$.

Appling linearly polarized laser parallel to the mirror plane does not break the mirror symmetry of the system, i.e., $\hat{M}_c \hat{H}(t) \hat{M}_c^{\dagger}=\hat{H}(t)$. The current perpendicular to the mirror plane meets $\hat{M}_c \mathbf{J}^a(\mathbf{K}, t) \hat{M}_c^{\dagger}=\mathbf{J}^{M_c^{-1} a}\left(M_c^{-1} \mathbf{K}, t\right)=-\mathbf{J}^a(\mathbf{K}, t)$. Here, the reverse current also comes from the reflection of the transition dipole moment. Since the currents on both sides of the mirror are opposite, it is concluded that when the driving light is parallel to the mirror plane, no harmonics can be generated perpendicular to the mirror plane (Relevant rules are presented in  Table I of main text).

\subsection*{C2. Rotational Symmetry}

\noindent Pure rotational symmetry usually exists in a two-dimensional plane, there are only 5 types of rotation axes in crystals $(1,2,3,4,6$-fold) due to periodic translational symmetry of lattices. Combining order-$N$ temporal operator $\hat{\tau}_N$ and $N$-fold rotational symmetry operator $\hat{C}_N$, we define a dynamical symmetry operation $\hat{R}_N=$ $\hat{C}_N \cdot \hat{\tau}_N$. When a laser field with $R_N$ symmetry acts on a crystal with $C_N$ symmetry (Considering the plane of electric field is always perpendicular to the $C_N$ axis), we have $\hat{R}_N \mathbf{k}(t)=C_N^{-1} \mathbf{K}+\mathbf{A}\left(\hat{\tau}_N t\right)=C_N^{-1} \mathbf{k}(t)$, this system can form a dynamical symmetry. The interband current transforms as
\begin{equation}
\begin{aligned}
&\hat{R}_N \mathbf{J}_{n m}^a(\mathbf{K}, t) \hat{R}_N^{\dagger}  =\mathbf{J}_{n m}^{C_N^{-1} a}\left(C_N^{-1} \mathbf{K}, \hat{\tau}_N t\right) \\
& =-i \varepsilon_{n m}\left(C_N^{-1} \mathbf{k}(t)\right) \mathbf{d}_{n m}^{C_N^{-1} a}\left(C_N^{-1} \mathbf{k}(t)\right) \rho_{n m}\left(C_N^{-1} \mathbf{K}, \hat{\tau}_N t\right) \\
& =-i \varepsilon_{n m}(\mathbf{k}(t)) \mathbf{d}_{n m}^a(\mathbf{k}(t)) e^{i\left[\phi_{n m}^{C_N^{-1} a}(\mathbf{k}(t))-\phi_{n m}^a(\mathbf{k}(t))\right]} \\
&\ \ \ \ \times\rho_{n m}(\mathbf{K}, t).
\end{aligned}
\end{equation}

Let's first consider the in-plane polarization. Since the rotational symmetry operator has two eigenvalues $e^{ \pm i \frac{2 \pi}{N}}$, the $\Delta \mathrm{TDP}$ should be $\phi_{n m}^{C_N^{-1} a}(\mathbf{k})-\phi_{n m}^a(\mathbf{k})=$ $\pm \frac{2 \pi}{N}$, denoting the rotation angle of dipoles under $\hat{C}_N$ [see Fig. 1(b) for the case of $N=3$ ]. Thus, integrating over the entire $\mathrm{BZ}$, we can obtain $\hat{R}_N \mathbf{J}_{n m}^a(t) \hat{R}_N^{\dagger}=$ $e^{ \pm i \frac{2 \pi}{N}} \mathbf{J}_{n m}^a(t)$. Using the interference form of the Fourier transform that $\hat{R}_N \mathbf{J}_{n m}^a(\omega) \hat{R}_N^{\dagger}=\mathbf{J}_{n m}^a(\omega)$, we know that the frequency of photons emitted perpendicular to the rotation axis can only be $\omega=(N l \pm 1) \omega_0, l \in \mathbb{N}$, which corresponds to co-rotating and counter-rotating photons relative to driving lasers, respectively. A same result can be obtained for the intraband current.

For the out-plane polarization, we have $\hat{R}_N \mathbf{J}^a(\mathbf{K}, t) \hat{R}_N^{\dagger}=\mathbf{J}^a(\mathbf{K}, t)$. Thus, the frequency of photons emitted parallel to the rotation axis can only be $\omega=N l \omega_0, l \in$ $\mathbb{N}$.

When the electric field is parallel to the rotation axis, the rotational symmetry of the system is always maintained. i.e., $\hat{C}_N \hat{H}(t) \hat{C}_N^{\dagger}=\hat{H}(t)$. The photocurrent perpendicular to the axis satisfies $\hat{C}_N \mathbf{J}^a(\mathbf{K}, t) \hat{C}_N^{\dagger}=\mathbf{J}^{C_N^{-1} a}\left(C_N^{-1} \mathbf{K}, t\right)=$ $e^{ \pm i \frac{2 \pi}{N}} \mathbf{J}^{C_N^{-1} a}\left(C_N^{-1} \mathbf{K}, t\right)$. Then we get $e^{ \pm i \frac{2 \pi}{N}}=1 \Rightarrow N=1$. In other words, when the electric field is always parallel to the $N$-fold $(N \geq 2)$ rotation axis, no current generates perpendicular to the axis (Relevant rules are presented in Table II of main text).

\section*{APPENDIX D: Derivation of Eqs. (12) and (17)}
\setcounter{equation}{0} 
\renewcommand{\theequation}{D\arabic{equation}}
We now derive the optical response induced by crystal time-reversal symmetry. For the proof of Eq. (12), considering time-reversal operation $\hat{T}$ on the Bloch state of electrons,
\begin{equation}
\hat{T}\left|u_{n, \mathbf{k}}\right\rangle=\left|u_{n,-\mathbf{k}}\right\rangle^* .
\end{equation}
Thus, for transition dipole matrix elements,
\begin{equation}
\begin{aligned}
\hat{T} \mathbf{d}_{n m}^a(\mathbf{k}) \hat{T}^{\dagger} & =\hat{T}\left\langle u_{n, \mathbf{k}}\left|i \partial_{\mathbf{k}_a}\right| u_{m, \mathbf{k}}\right) \hat{T}^{\dagger} \\
& =-i\left\langle\partial_{-\mathbf{k}_a} u_{m,-\mathbf{k}} \mid u_{n,-\mathbf{k}}\right\rangle \\
& =i\left\langle u_{m,-\mathbf{k}}\left|\partial_{-\mathbf{k}_a}\right| u_{n,-\mathbf{k}}\right\rangle \\
& =\mathbf{d}_{m n}^a(-\mathbf{k}) .
\end{aligned}
\end{equation}
If the system has time-reversal symmetry (i.e., $\hat{T} \hat{H}=\hat{H} \hat{T}$ ), the Bloch wave is the eigenstate of $\hat{T}$ as well. Due to the antiunitarity of $\hat{T}$, its eigenvalues are complex numbers of modulo 1 :
\begin{equation}
\hat{T}\left|u_{n, \mathbf{k}}\right\rangle=e^{i \phi_{\mathbf{k}}}\left|u_{n, \mathbf{k}}\right\rangle.
\end{equation}
Therefore, we have $\left|u_{n,-\mathbf{k}}\right\rangle^*=e^{i \phi_{\mathbf{k}}}\left|u_{n, \mathbf{k}}\right\rangle$. Then,
\begin{equation}
\begin{aligned}
\mathbf{d}_{m n}^a(-\mathbf{k}) & =-i\left\langle\partial_{-\mathbf{k}_a} u_{m,-\mathbf{k}} \mid u_{n,-\mathbf{k}}\right\rangle \\
& =i\left\langle u_{n, \mathbf{k}}\left|\partial \mathbf{k}_a\right| u_{m, \mathbf{k}}\right\rangle \\
& =\mathbf{d}_{n m}^a(\mathbf{k}).
\end{aligned}
\end{equation}
Similarly, we can obtain the constraints of time-reversal symmetry on the band dispersion and the shift vector, respectively:
\begin{equation}
\begin{aligned}
\hat{T} \varepsilon_n(\mathbf{k}) \hat{T}^{\dagger} & =\varepsilon_n(-\mathbf{k})=\varepsilon_n(\mathbf{k}),\\
\hat{T} \mathbf{R}_{n m}^{a, b}(\mathbf{k}) \hat{T}^{\dagger} & =\mathbf{R}_{n m}^{a, b}(-\mathbf{k})=\mathbf{R}_{n m}^{a, b}(\mathbf{k}) .
\end{aligned}
\end{equation}
Similar to the case of inversion symmetry, we define an operator $\hat{U}$ that performs $\hat{T}$ on the crystals, and order-2 temporal operator $\hat{\tau}_2$ on the time. The time-dependent quasi momentum has $\hat{U} \mathbf{k}(t)=-\mathbf{k}(t)$. The dynamical phase, shift phase and $\Delta$TDP transform as
\begin{equation}
\begin{aligned}
\hat{U} S_{\mathrm {dyn}}\left(\mathbf{K}, t, t^{\prime}\right) \hat{U}^{\dagger} & =S_{\mathrm {dyn}}\left(-\mathbf{K}, \hat{\tau}_2 t, \hat{\tau}_2 t^{\prime}\right)\\
&=S_{\mathrm {dyn}}\left(\mathbf{K}, t, t^{\prime}\right), \\
\hat{U} S_{\mathrm {shift}}\left(\mathbf{K}, t, t^{\prime}\right)
\hat{U}^{\dagger} 
& =S_{\mathrm {shift}}\left(-\mathbf{K}, \hat{\tau}_2 t, \hat{\tau}_2 t^{\prime}\right)\\
&=-S_{\mathrm {shift}}\left(\mathbf{K}, t, t^{\prime}\right), \\
\hat{U} S_{\Delta \mathrm{TDP}}(\mathbf{k}(t)) \hat{U}^{\dagger} & =S_{\Delta \mathrm{TDP}}(-\mathbf{k}(t))\\
&=-S_{\Delta \mathrm{TDP}}(\mathbf{k}(t)) .
\end{aligned}
\end{equation}
Combining the point-group symmetry, 
\begin{equation}
\begin{aligned}
\hat{P} S_{\mathrm {shift}}\left(\mathbf{K}, t, t^{\prime}\right) \hat{P}^{\dagger} & =S_{\mathrm {shift}}\left(-\mathbf{K}, \hat{\tau}_2 t, \hat{\tau}_2 t^{\prime}\right)\\
&=S_{\mathrm {shift}}\left(\mathbf{K}, t, t^{\prime}\right), 
\end{aligned}
\end{equation}
we know that the $S_{\mathrm {shift }}$ vanish in crystals that both time-reversal symmetry and inversion symmetry are satisfied. Therefore, it could be reasonable to consider only the dynamical phase at this time. Of course, when considering the harmonics that are not collinear with the laser field, $\Delta \mathrm{TDP}$ exactly cannot be ignored.

The interband current under $\hat{U}$ takes 
\begin{widetext}
\begin{equation}
\begin{aligned}
\hat{U} \mathbf{J}_{n m}^a(\mathbf{K}, t) \hat{U}^{\dagger}= & \mathbf{J}_{n m}^a\left(-\mathbf{K}, \hat{\tau}_2 t\right) \\
= & -\int_{-\infty}^{\hat{\tau}_2 t} d \hat{\tau}_2 t^{\prime} \varepsilon_{n m}(-\mathbf{k}(t))\left|\mathbf{d}_{n m}^a(-\mathbf{k}(t))\right|\left[\mathbf{E}^b\left(\hat{\tau}_2 t^{\prime}\right) \cdot\left|\mathbf{d}_{m n}^b\left(-\mathbf{k}\left(t^{\prime}\right)\right)\right|\right] \\
& \times f_{n m}\left(-\mathbf{K}, \hat{\tau}_2 t\right) e^{-i\left[S_{\mathrm{dyn}}\left(-\mathbf{K}, \hat{\tau}_2 t, \hat{\tau}_2 t^{\prime}\right)+S_{\mathrm{shift}}\left(-\mathbf{K}, \hat{\tau}_2 t, \hat{\tau}_2 t^{\prime}\right)+S_{\Delta \mathrm{TDP}}(-\mathbf{k}(t))\right]} \\
= & \int_{-\infty}^t d t^{\prime} \varepsilon_{n m}(\mathbf{k}(t))\left|\mathbf{d}_{m n}^a(\mathbf{k}(t))\right|\left[\mathbf{E}^b\left(t^{\prime}\right) \cdot\left|\mathbf{d}_{n m}^b\left(\mathbf{k}\left(t^{\prime}\right)\right)\right|\right] \\
& \times f_{n m}(\mathbf{K}, t) e^{-i\left[S_{\mathrm{dyn}}\left(\mathbf{K}, t, t^{\prime}\right)-S_{\mathrm{shift}}\left(\mathbf{K}, t, t^{\prime}\right)-S_{\Delta \mathrm{TDP}}(\mathbf{k}(t))\right]} \\
= & -\mathbf{J}_{n m}^a(\mathbf{K}, t) e^{2 i\left[S_{\mathrm{shift}}\left(\mathbf{K}, t, t^{\prime}\right)+S_{\Delta \mathrm{TDP}}(\mathbf{k}(t))\right]}.
\end{aligned}
\end{equation}
\end{widetext}

For the proof of Eq. (17),
\begin{widetext}
\begin{equation}
\begin{aligned}
\hat{T} \mathbf{J}_{n m, \sigma_{+}}^a(\mathbf{K}, t) \hat{T}^{\dagger}= & \mathbf{J}_{n m, \sigma_{-}}^a(-\mathbf{K},-t) \\
= & -\int_{+\infty}^{-t} d t^{\prime} \varepsilon_{n m}\left(-\mathbf{K}+\mathbf{A}_{\sigma_{-}}(-t)\right)\left|\mathbf{d}_{n m}^a\left(-\mathbf{K}+\mathbf{A}_{\sigma_{-}}(-t)\right)\right| \\
& \times {\left[\mathbf{E}_{\sigma_{-}}^b\left(t^{\prime}\right) \cdot\left|\mathbf{d}_{m n}^b\left(-\mathbf{K}+\mathbf{A}_{\sigma_{-}}\left(t^{\prime}\right)\right)\right|\right] f_{n m}(-\mathbf{K},-t) } \\
& \times e^{-i\left[S_{\mathrm{dyn}}\left(-\mathbf{K}, \mathbf{A}_{\sigma_{-}},-t, t^{\prime}\right)+S_{\mathrm{shift}}\left(-\mathbf{K}, \mathbf{A}_{\sigma_{-}}-t, t^{\prime}\right)+S_{\Delta \mathrm{TDP}}\left(-\mathbf{K}+\mathbf{A}_{\sigma_{-}}(-t)\right)\right]} \\
= & \int_{-\infty}^t d t^{\prime} \varepsilon_{n m}\left(\mathbf{K}+\mathbf{A}_{\sigma_{+}}(t)\right)\left|\mathbf{d}_{m n}^a\left(\mathbf{K}+\mathbf{A}_{\sigma_{+}}(t)\right)\right| \\
& \times {\left[\mathbf{E}_{\sigma_{+}}^b\left(t^{\prime}\right) \cdot\left|\mathbf{d}_{n m}^b\left(\mathbf{K}+\mathbf{A}_{\sigma_{+}}\left(t^{\prime}\right)\right)\right|\right] f_{n m}(-\mathbf{K},-t) } \\
& \times e^{-i\left[-S_{\mathrm{dyn}}\left(\mathbf{K}, \mathbf{A}_{\sigma_{+}}, t, t^{\prime}\right)-S_{\mathrm{shift}}\left(\mathbf{K}, \mathbf{A}_{\sigma_{+}}, t, t^{\prime}\right)-S_{\Delta \mathrm{TDP}}\left(\mathbf{K}+\mathbf{A}_{\sigma_{+}}(t)\right)\right]} \\
= & -\mathbf{J}_{n m, \sigma_{+}}^{a, *}(\mathbf{K}, t) \frac{f_{n m}(-\mathbf{K},-t)}{f_{n m}(\mathbf{K}, t)} .
\end{aligned}
\end{equation}
\end{widetext}
Here, the initial right-hand helically polarized laser $\left(\sigma_{+}\right)$is changed to left-hand helically polarized laser $\left(\sigma_{-}\right)$under the time-reversal transformation, which have $\mathbf{E}_{\sigma_{+}}(t)=\mathbf{E}_{\sigma_{-}}(-t)$, and $\mathbf{A}_{\sigma_{+}}(t)=-\mathbf{A}_{\sigma_{-}}(-t)$. In the second step, we have assumed that the pulse envelope is infinite, then the temporal integral from $-\infty$ is identical with that from $+\infty$. We use the transformation relations shown in Eqs. (14-16) in the third step. A derivation process for the intraband current is similar.

\bibliography{bibfile}

\end{document}